\documentclass[twocolumn,preprintnumbers,amsmath,amssymb,superscriptaddress]{revtex4-2}

\usepackage[utf8]{inputenc}
\usepackage{amssymb}
\usepackage{amsmath}
\usepackage{amsfonts}
\usepackage{graphicx}
\usepackage{newunicodechar}
\newunicodechar{₅}{$_5$}
\newunicodechar{⁺}{\textsuperscript{+}}

\usepackage{cancel}

\usepackage{newfloat}
\DeclareFloatingEnvironment[
  listname = {Case studies} ,
  name = {CASE STUDY}https://www.overleaf.com/project/639065ef54e709d395b47d76
]{casestudy}

\usepackage[normalem]{ulem}
\usepackage{color}
\usepackage{xcolor}

\usepackage{braket}

\begin{document}
\title{Dynamical Valence Fluctuations and Magnetic Anisotropy in Sm(Co,Cu)$_5$}
\author{Alexander B. Shick}
\email{shick@fzu.cz}
\affiliation{Institute of Physics, the Czech Academy of Sciences,  Na Slovance 2, CZ-18221 Prague, Czech Republic}
\author{Alexander V. Andreev}
\affiliation{Institute of Physics, the Czech Academy of Sciences,  Na Slovance 2, CZ-18221 Prague, Czech Republic}
\author{Evgenia A. Tereshina-Chitrova}
\affiliation{Institute of Physics, the Czech Academy of Sciences,  Na Slovance 2, CZ-18221 Prague, Czech Republic}

\date{\today}

\begin{abstract}

The magnetic properties of Sm-based intermetallics are governed by a delicate interplay between localized $4f$ states, exchange interactions, and valence fluctuations, yet the role of dynamical $4f$ correlations remains poorly understood. Here we investigate the electronic structure and magnetic anisotropy of SmCo$_5$ and SmCu$_5$ using self-consistent DFT+$U$(ED), which explicitly treats dynamical fluctuations of the Sm $4f$ shell. We find that SmCo$_5$ exhibits pronounced Sm$^{2+}$--Sm$^{3+}$ valence fluctuations, with the $4f$ ground state dominated by $f^5$ and $f^6$ configurations. The resulting dynamical $4f$ correlations yield near compensation of the Sm spin and orbital moments and provide a quantitative description of the electronic structure, magnetic moments, and large uniaxial magnetic anisotropy of SmCo$_5$. Upon replacing Co with Cu, the balance shifts toward the $f^6$ configuration, with substantial changes in the Sm spin and orbital moments. In SmCu$_5$, this evolution of the $4f$ configuration is accompanied by a pronounced reduction of both the total magnetic moment and magnetic anisotropy. These results demonstrate that dynamical $4f$ valence fluctuations are essential for understanding the contrasting magnetic properties of SmCo$_5$ and SmCu$_5$ and provide a microscopic framework for describing magnetic anisotropy upon Co to Cu substitution in Sm(Co,Cu)$_5$.

\end{abstract}

\maketitle

The rare-earth intermetallic compound SmCo$_5$ {formed the basis of the first generation of} high-performance rare-earth permanent magnets, combining high saturation magnetization and a high Curie temperature with excellent uniaxial magnetocrystalline anisotropy energy (MAE) of about 9.2 meV/f.u. (17.2 MJm$^{-3}$)~\cite{coey2011}. {Despite their high cost and brittleness, SmCo$_5$-based magnets remain technologically important because of their excellent thermal stability and resistance to demagnetization at elevated temperatures~\cite{gutfleisch2011}. 

\begin{figure}[!hb]
\centerline{\hspace*{2cm} \includegraphics[width=3.5in]{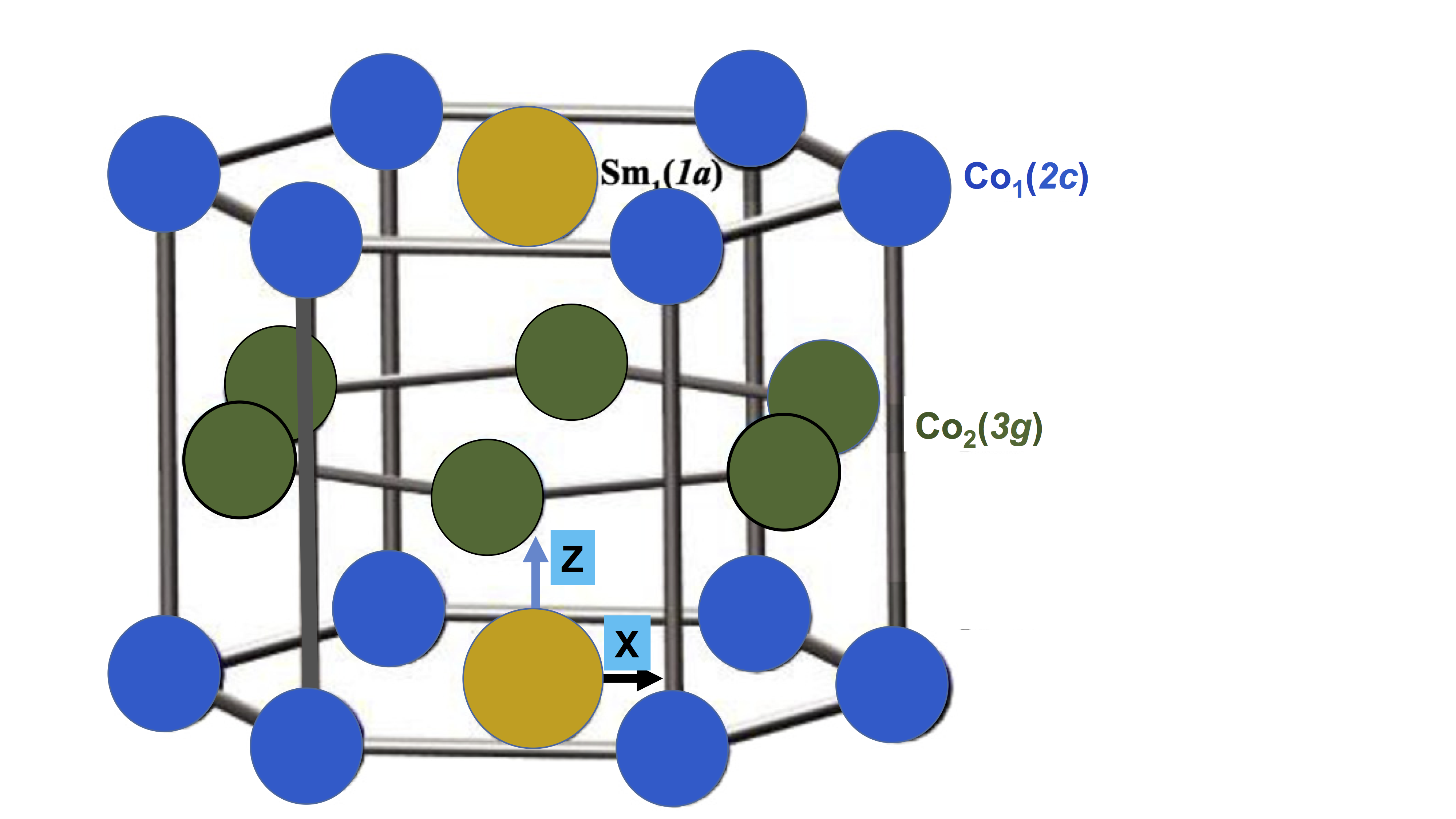}}
\caption{The hexagonal P6/mmm SmCo$_5$ crystal structure.}
\label{fig:1}
\end{figure}

Sm-Co alloys have also attracted interest as high-anisotropy media for ultrahigh-density magnetic recording and heat-assisted magnetic recording (HAMR)~\cite{sayama2005smco}. Recent studies of Sm-Co-based thin films have demonstrated high coercivity together with enhanced remanence through compositional modification, and have identified Sm(Co,Cu)$_5$-based films as promising candidates for HAMR applications~\cite{zheng2025}. The MAE of SmCo$_5$ provides high thermal stability at reduced magnetic dimensions, making it an interesting alternative to the $L1_0$-FePt media that currently dominate HAMR development~\cite{weller2013l10}. In contrast to FePt, however, Sm-Co films have received considerably less attention as practical granular HAMR media, and their large anisotropy field and microstructural complexity remain important challenges for high-density recording. Patterning and compositional engineering of Sm--Co films therefore provide possible routes toward exploiting their large anisotropy while improving their writability and magnetic isolation.

SmCo$_5$ {crystallizes} in a hexagonal CaCu$_5$-type crystal structure (space group $P6/mmm$) with lattice parameters $a$ = 4.97~\AA ~ and $c$ = 3.98~\AA, and {the hexagonal \(c\) axis as the easy magnetization direction}~\cite{andreev1995}. The Co atoms occupy two distinct crystallographic sites: 2c and 3g, while the Sm atoms occupy the 1a site (see Fig.~\ref{fig:1}). 

Soon after the discovery of SmCo$_5$, it was recognized that the crystal-field (CF) interaction of the Sm$^{3+}$ ion in the hexagonal crystal field is the main reason for the high anisotropy \cite{buschow1977}. 
Crystal‑field theory~\cite{jensen1991rare} predicts for Sm³⁺ in SmCo$_5$ a large orbital contribution ($\sim$ 4 $\mu_B$) that is almost cancelled by an opposite spin contribution ($\sim$ ‑3.5 $\mu_B$), leaving a tiny net Sm moment of roughly 0.3–0.4 $\mu_B$ at low temperature ~\cite{coey2002permanent}. This tiny but highly axial Sm moment is what supplies the extraordinary uniaxial anisotropy of the compound, even though the Co sub-lattice dominates the bulk magnetisation.

The CF theory experiences some difficulties when applied to SmCo$_5$ due to the breakdown of the simple separation between atomic multiplet physics and CF perturbation. Within the CF model, the localized Sm $4f$ electrons are first described by the atomic Coulomb and spin-orbit interactions, which give rise to well-defined $LSJ$ multiplets, and the CF is subsequently treated as a perturbation that lifts the degeneracy of these multiplets~\cite{taylor1972physics}. For most rare-earth ions, the separation between the ground-state and excited $J$ multiplets is sufficiently large that the crystal-field Hamiltonian can be projected onto the lowest $J$ manifold and expressed in terms of Stevens operators. Sm$^{3+}$ is unusual in this respect because its $J=5/2$ ground multiplet is separated from the first excited $J=7/2$ multiplet by only about 0.12 eV. As a result, the CF can induce appreciable admixture of excited $J$ states, so that a quantitative description of the magnetic properties of SmCo$_5$ may require a CF treatment beyond the single-$J$ approximation, with the $J=5/2$ and $J=7/2$ manifolds included explicitly.

The standard rare-earth crystal-field model provides a phenomenological description of the Sm $4f$ states, but does not constitute a genuine first-principles theory. Several attempts have therefore been made to describe SmCo$_5$ from first principles using density functional theory (DFT) supplemented by a Hubbard interaction (DFT+$U$)~\cite{larson2003,landa2018,das2019} (see the Supplemental Material (SM)~\cite{SM}).

The DFT+$U$ calculations of Larson and Mazin~\cite{larson2003} yielded a total magnetic moment and uniaxial MAE in reasonable agreement with experiment~\cite{coey2011,omari2002}. Subsequent DFT+$U$ calculations by Landa {\em et al.}~\cite{landa2018}, however, obtained a negative MAE, at variance with the well-established experimental uniaxial anisotropy. Furthermore, Landa {\em et al.} demonstrated that the solution reported in Ref.~\cite{larson2003} corresponds to a parallel alignment of the Sm and Co spin moments, $M_S$, whereas the ground state of SmCo$_5$ is characterized by the experimentally established antiparallel alignment of the rare-earth and transition-metal spins~\cite{campbell1972indirect,brooks1989}. This parallel-spin configuration is a metastable state lying a few mRy above the antiparallel ground state.

An alternative description of SmCo$_5$ was proposed by Das {\em et al.}~\cite{das2019} within DFT+$U$. Their calculations predict a substantially quenched Sm orbital moment and an electronic structure that deviates from the conventional Hund's-rule description of the Sm $4f$ shell. The authors further emphasized that this picture differs from both the conventional crystal-field (CF) description and the available experimental evidence.

In contrast, self-interaction-corrected DFT (SIC-DFT) calculations by Patrick and Staunton~\cite{patrick2019} recover the antiparallel alignment of the Sm and Co spin moments in the ground state. They also obtain the total magnetic moment and MAE (shown in Table~S1 of SM~\cite{SM}) in a reasonable agreement with experiment. Nevertheless, the SIC-DFT results do not exhibit the substantial compensation between the Sm spin and orbital moments expected from the CF picture and observed in polarized-neutron diffraction experiments~\cite{givord1979temperature}.

Attempts to go beyond the static mean-field treatment of the Sm $4f$ shell in SmCo$_5$ have also been made using the ''quasi-atomic'' Hubbard-I approximation (HIA), in which the hybridization of the $4f$ states with the valence bands is neglected. Granas {\em et al.}~\cite{graanas2012charge} employed HIA within dynamical mean-field theory (DMFT) and obtained antiparallel Sm spin and orbital moments, with their magnitudes nearly compensating each other. The Co spin moment is antiparallel to the Sm spin moment, yielding a total magnetic moment of approximately $8~\mu_B$/f.u., in reasonable agreement with experiment. Ref.~\cite {graanas2012charge} reported no magnetic-anisotropy-energy (MAE) calculations. An independent DFT+HIA study~\cite{shick2021electronic} likewise found compensated Sm spin and orbital moments and antiparallel alignment of the Co and Sm spin moments. Also, the calculated MAE agreed well with experiment.

In this work, the DFT+$U$(ED) approach~\cite{shick2021,shick2026magnetic}, which combines DFT+$U$ with the exact diagonalization (ED) of the multiorbital Anderson impurity model (AIM)~\cite{mahan2013}  for the 4$f$ shell, is used to investigate the remarkable spin and orbital magnetism, and MAE in SmCo$_5$. The electron-electron interactions in the 4$f$ shell are treated explicitly using the exact diagonalization of the AIM, whereas the $s$, $p$, and $d$ shells are treated within DFT. Further details of the DFT+$U$(ED) implementation can be found in~\cite{shick2021,shick2026magnetic,SM}.

For the DFT+$U$ calculations, an in-house implementation~\cite{shick2001} of the full-potential linearized augmented plane-wave (FP-LAPW) method was employed, including both scalar-relativistic and spin-orbit coupling effects. The experimental lattice constants~\cite{andreev1995} were used for the calculations. The conventional local spin-density approximation (LSDA)
is adopted, which is expected to be valid for itinerant metallic systems. The radii of the atomic muffin-tin (MT) spheres are set 
to 2.75~a.u. for the Sm atom, and 2.20~a.u. for Co atoms. The parameter $R_{Co} \times K_{max}=7.7$ defines the basis set size and the Brillouin zone (BZ) was sampled with 2535 $k$~points.   
The Slater integrals  $F_0=6.87$~eV, and $F_2=9.06$ eV, $F_4=6.05$ eV, and $F_6=$4.45 eV were chosen which correspond to the Hubbard parameters 
$U = 6.87$~eV and $J = 0.76$~eV, in the range of commonly accepted values for
 the rare earth elements ~\cite{vdmarel1988}.
 \begin{table*}[!htbp]
\caption{The spin and orbital magnetic moments $\mu_B$, aligned along [001] crystal direction.} 
\centering
\begin{tabular}{llcccccccccc}
\hline
\multicolumn{1}{l}{SmCo$_5$} & \multicolumn{2}{c}{Sm-$M^S_{4f}/M^L_{4f}/M^J_{4f}$} & \multicolumn{2}{c}{Sm(1a)-$M^S/M^L/M^J$} &\multicolumn{2}{c}{Co(2c)-$M^S/M^L/M^J$} &\multicolumn{2}{c}{Co(3g)-$M^S/M^L/M^J$} & Total M/f.u.  \\
\hline
\multicolumn{1}{l}{DFT+$U$(ED)} & \multicolumn{2}{c}{-4.11/3.99/-0.12} & \multicolumn{2}{c}{-4.39/4.01/-0.38} & \multicolumn{2}{c}{1.47/0.10/1.57}& \multicolumn{2}{c}{1.52/0.07/1.59} & 7.06 \\
\hline
\multicolumn{1}{l}{Experiment}    &&&\multicolumn{2}{c}{-0.38~\cite{givord1979temperature}}&\multicolumn{2}{c}{1.86~\cite{givord1979temperature}}
&\multicolumn{2}{c}{1.75~\cite{givord1979temperature}}&\multicolumn{2}{c}{7.8~\cite{coey2010magnetism}, 8.9~\cite{tie1991magnetic}}\\
\hline
\multicolumn{1}{l}{SmCu$_5$} & \multicolumn{2}{c}{Sm-$M^S_{4f}/M^L_{4f}/M^J_{4f}$} & \multicolumn{2}{c}{Sm(1a)-$M^S/M^L/M^J$} &\multicolumn{2}{c}{Cu(2c)-$M^S/M^L/M^J$} &\multicolumn{2}{c}{Cu(3g)-$M^S/M^L/M^J$} & Total M/f.u.  \\
\hline
\multicolumn{1}{l}{DFT+$U$(ED)} & \multicolumn{2}{c}{-3.34/1.85/-1.49} & \multicolumn{2}{c}{-3.46/1.85/-1.61} & \multicolumn{2}{c}{0/0/0}& \multicolumn{2}{c}{-0.01/0.0/-0.01} & -1.87 \\
\hline
\end{tabular}
\label{tab:1}
\end{table*}

The calculated spin ($M_S$), orbital ($M_L$), and total ($M_J$) magnetic moments of the Sm and Co sites, together with the total magnetic moment $M$ per formula unit, are summarized in Table~\ref{tab:1}. Within the DFT+$U$(ED) approach, the Sm $4f$ shell has an occupation of $n_f=5.25$, with spin and orbital moments of $M^{S}{4f}=-4.105~\mu_B$ and $M^{L}{4f}=3.99~\mu_B$, respectively. The Sm spin moment is antiparallel to the Co spin moments on both the Co(2c) and Co(3g) sublattices. The resulting Sm total moment, $M_J=-0.38~\mu_B$~\cite{givord1979temperature}, is in quantitative agreement with the experimental value. The calculated Co moments on both crystallographic sites are somewhat smaller than the corresponding experimental values, resulting in a total magnetic moment below experiment. As discussed below, this discrepancy can be traced to the reduced Co contribution to the magnetization.
\begin{figure}[!htbp]
\centerline{\hspace*{0cm} \includegraphics[width=4in]{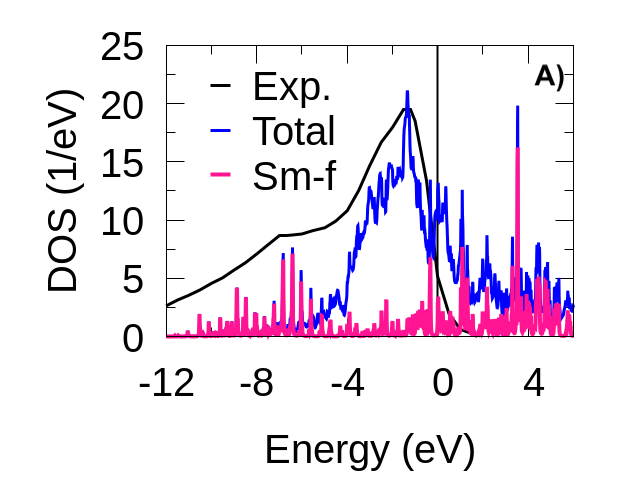}}
\centerline{\hspace*{0cm} \includegraphics[width=4in]{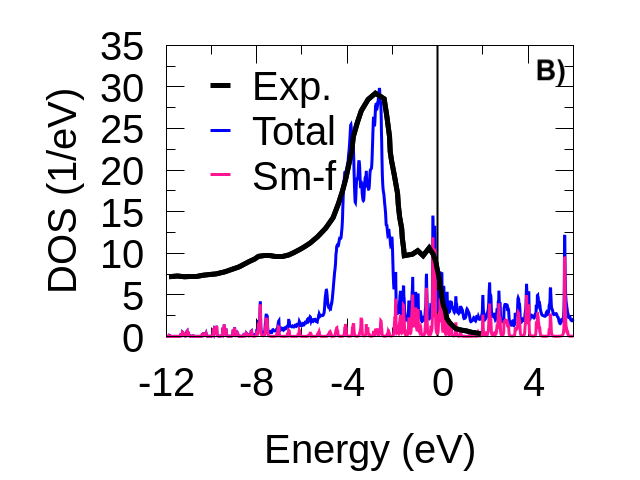}}
	\caption{Sm 4$f$-projected DOS and non-$f$ itinerant DOS per unit cell for (A) SmCo$_5$ together with experimental valence PES data~\cite{cuthill1974}, and (B)  SmCu$_5$ with experimental PES spectrum of Sm/Cu films~\cite{andersen1988mixed}.}
	\label{fig:2}
\end{figure}

A direct validation of the calculated electronic structure is provided by its agreement with photoemission spectroscopy (PES). Figure~\ref{fig:2}A compares the Sm $f$-projected and total densities of states (DOS) obtained within DFT+$U$(ED) with the experimental PES spectrum~\cite{cuthill1974}. In the quasi-atomic DFT+HIA approach, the occupied Sm $4f$ states are concentrated in a narrow structure at approximately 5.5~eV below the Fermi level, $E_F$, shown in Fig. 1 of Ref.~\cite{shick2021electronic}. Inclusion of the $4f$--valence-electron hybridization within DFT+$U$(ED) substantially redistributes the Sm $f$ spectral weight: the predominantly localized occupied peaks shift to larger binding energies and lose intensity, while additional $4f$ spectral weight emerges in the binding-energy range from $-3$ to $0$~eV. The Co $d$ states form the dominant itinerant contribution over the $-4$ to $0$~eV range. The resulting combined Sm $f$ and Co $d$ spectral features are in good agreement with the experimental PES, demonstrating the importance of hybridization for a quantitative description of the SmCo$_5$ electronic structure.

For comparison, SIC-DFT places the occupied Sm $4f$ states at approximately 11~eV below $E_F$~\cite{patrick2018}, substantially deeper than observed experimentally and in marked contrast to the DFT+$U$(ED) results.

Since there is no direct experimental evidence of the presence of the 4$f$ states near $E_F$, which could be provided by resonant PES near 3$d$ $\rightarrow$ 4$f$ transition, we desided to examine the modification of the 3$d$ states, and performed the calculations of the SmCu$_5$.
SmCu$_5$ stabilizes in the same hexagonal CaCu$_5$-type crystal structure, and orders magnetically at $\approx$ 9 K~\cite{svoboda1990magnetic}. In the SmCu$_5$ calculations, we used the experimental lattice constants~\cite{buschow1971composition} and assumed the ferromagnetic order
of the Sm atoms.

The calculated $M_S$, $M_L$, and $M_J$ magnetic moments of the Sm and Cu sites, together with the total magnetic moment $M$ per formula unit, are summarized in Table~\ref{tab:1}. The Sm $4f$ shell has an occupation of $n_f=5.64$, with spin and orbital moments of $M^{S}_{4f}=-3.34~\mu_B$ and $M^{L}_{4f}=1.85~\mu_B$, respectively. The Cu(2c) and Cu(3g) sites carry practically zero spin and orbital moments. The resulting total Sm moment is $M_J=-1.87~\mu_B$.

Figure~\ref{fig:2}B compares the Sm $f$-projected and total densities of states (DOS) obtained within DFT+$U$(ED) with the experimental PES spectrum of Sm/Cu films~\cite{andersen1988mixed}. The Cu $3d$ states provide the dominant contribution to the total DOS in the $-5$ to $-2$~eV binding-energy range, in good agreement with experiment. The calculated Sm $4f$ spectral weight is distributed over a broad energy range, with pronounced weight between $-2$ and $0$~eV and an additional peak near $-8$~eV, both of which are also observed in the experimental PES spectrum. Overall, the calculated DOS of bulk SmCu$_5$ shows surprisingly good semi-quantitative agreement with the PES data for Sm/Cu films~\cite{andersen1988mixed}. 

Within the DFT+$U$(ED) framework, the calculated Sm $4f$-state occupation in SmCo$5$ is $n_f=5.25$, corresponding to a many-body ground state (GS) of the Anderson impurity model. The probabilities $P_{f^n}$ of finding the Sm ion in atomic eigenstates with integer $4f$ occupations are $P_{f^5}=0.74$, $P_{f^6}=0.25$, and $P_{f^7}=0.01$. Thus, the GS is predominantly a superposition of $f^5$ and $f^6$ configurations, indicating intermediate valence and Sm$^{2+}$--Sm$^{3+}$ valence fluctuations in SmCo$_5$.

For SmCu$5$, the calculated occupation increases to $n_f=5.64$, with $P_{f^5}=0.37$, $P_{f^6}=0.62$, and $P_{f^7}=0.01$. Compared with SmCo$_5$, the reduced $f^5$ weight and enhanced $f^6$ weight indicate a substantially stronger contribution of the Sm$^{2+}$ configuration in SmCu$_5$. This redistribution of the configuration weights is also reflected in the calculated Sm $f$-projected DOS. In SmCo$5$, the spectral feature at $\sim -7$~eV predominantly corresponds to the $f^5\rightarrow f^4$ removal transition, whereas the spectral weight near $E_F$ is associated with the $f^6\rightarrow f^5$ transition. In SmCu$5$, the corresponding transitions give rise to features near $\sim -8$~eV and close to $E_F$, respectively. The changes in the relative intensities of these spectral features follow the redistribution of the $P_{f^5}$ and $P_{f^6}$ configuration weights, providing a direct connection between the many-body configuration statistics and the observed $f$-electron spectral structure.

Now, we turn to the magnetic anisotropy properties of SmCo$_5$ and SmCu$_5$.
The uniaxial MAE is evaluated following the two-sublattice model previously applied to 3d-4f intermetallics~\cite{boust2022and}. In this approach, the magnetic system is separated into a 3d transition-metal sublattice and a localized Sm (4f) sublattice. The Sm contribution is obtained by considering a single Sm ion embedded in the crystal environment and subject to the exchange field generated by the Co sublattice. From the self-consistent DFT+$U$(ED) solution, we extract all parameters entering Eq.~(S1), including the SOC coupling constant ($\xi$), the crystal-field matrix $\Delta_{CF}$, and the exchange field ($\Delta_{EX}$). The Sm-MAE contribution is then evaluated from the difference between the ground-state energies of Eq.~(S1) for $\Delta_{EX}$ oriented along the in-plane [110] and out-of-plane [001] directions (see Fig.~\ref{fig:1}). The Co contribution is evaluated separately by replacing Sm with Y and calculating the MAE of YCo5 as the difference in total energy between magnetization along the the in-plane [110] and out-of-plane [001] directions within DFT or DFT+$U$. The total MAE is subsequently obtained as the sum of the Sm and Co contributions.

The calculated uniaxial MAE values are summarized in Table~\ref{tab:2}. When DFT+$U$ is applied to the Co $3d$ states, with $U=1.9$~eV and $J=0.8$~eV~\cite{nguyen2018magnetocrystalline}, the Co contribution to the MAE increases substantially. The resulting MAE of SmCo$_5$ is in very good agreement with the low-temperature bulk experimental values~\cite{omari2002,ermolenko1976} and somewhat larger than the room-temperature value~\cite{coey2010magnetism}. 

Furthermore, applying DFT+$U$ to the Co $3d$ shell introduces $3d$ orbital polarization through the Coulomb interaction $U$ and exchange interaction $J$, resulting in a pronounced enhancement of the Co orbital moments $M_L$ and, consequently, of the total magnetic moment $M$~\cite{SM}. The calculated total moment of $9.27~\mu_B$/f.u. is in good agreement with the available low-temperature experimental data~\cite{tie1991magnetic}.

In SmCu$_5$, the Sm contribution to the MAE is strongly reduced, accompanied by a substantial reduction in the total magnetic moment. Since the Cu $3d$ states carry essentially negligible spin and orbital moments, as shown in Table~\ref{tab:1}, 
they do not contribute to the MAE.
 
\begin{table}[!h]
\caption{Total and element-specific MAE contributions (meV) calculated within the two-sublattice model, compared with experimental data for SmCo$_5$ and SmCu$_5$
}
\centering
\begin{tabular}{cccccccccc}
\hline
  SmCo$_5$           & Sm-MAE                 &  Co-MAE & MAE \\ 
\hline
DFT+$U$(ED)            & 12.39                  &  0.27   & 12.66  \\
{DFT+$U$(ED) + }       & 12.39                  & 1.55    & 13.94 \\
DFT+$U$ on 3$d$-Co   &                      \\
\hline
Experiment&           &                   & 9.2~\cite{coey2010magnetism}  \\ 
    &           &         & 13.1~\cite{omari2002},17.4~\cite{ermolenko1976}  \\
\hline
  SmCu$_5$    & Sm-MAE       & Cu-MAE            &  MAE    \\ 
\hline
{DFT+$U$(ED)} & 2.64           &  0                &  2.64   \\ \hline
\end{tabular}
\label{tab:2}
\end{table}

In conclusion, our results establish the importance of dynamical valence fluctuations for the electronic and magnetic properties of SmCo$_5$ and SmCu$_5$. In SmCo$_5$, the Sm $4f$ shell is characterized by a ground-state superposition of predominantly $f^5$ ($P_{f^5}=0.74$) and $f^6$ ($P_{f^6}=0.25$) configurations, demonstrating pronounced Sm$^{2+}$--Sm$^{3+}$ valence fluctuations. Substitution of Co by Cu substantially shifts this balance toward the $f^6$ configuration, with $P_{f^6}=0.62$ and $P_{f^5}=0.37$, reflecting the reduced exchange field acting on the Sm $4f$ shell. Accounting for these dynamical fluctuations within DFT+$U$(ED) yields a nearly compensated spin and orbital contributions of the Sm $4f$ shell in SmCo$_5$ and provides a consistent quantitative description of its electronic structure, magnetic moments, and large uniaxial MAE. The same framework captures the pronounced reduction of the magnetic anisotropy and total magnetic moment upon Co-to-Cu substitution in SmCu$_5$. Our results thus highlight the essential role of dynamical $4f$ correlations in Sm(Co,Cu)$_5$ and provide a microscopic basis for understanding their magnetic properties, with implications for the development of Sm(Co,Cu)$_5$-based media for ultrahigh-density magnetic recording.

\section*{Acknowledgements}
Partial support provided by the Czech Science Foundation (GACR) Grant No. 24-11992S is acknowledged.
This work was partially co-funded by the European Union and the Czech Ministry of Education, Youth and Sports (Project TERAFIT -
CZ.02.01.01/00/22$_{-}$008/0004594).
A.V.A. and E.A.T. acknowledge partial support provided by the Czech Science Foundation (GACR) Grant No. 26-20860S.

\bibliography{refs,refs2}

\pagebreak


\end{document}